# Programming electronic states in conductive metal–organic frameworks through crystallization-enabled proton management

Hao Chen[1,2], Tongyang Zhao[2], Weishan Li[2], Jinkun Guo[2], Jia-Xiang Zhang[2], Ze-Fan Yao[3], Maojun Zheng[1] and Jin-Hu Dou[2]*

[1]Key Laboratory of Artificial Structure and Quantum Control, Ministry of Education, School of Physics and Astronomy, Shanghai Jiao Tong University, Shanghai, 200240, China.

[2]National Key Laboratory of Advanced Micro and Nano Manufacture Technology; Key Laboratory of Polymer Chemistry and Physics of Ministry of Education, School of Materials Science and Engineering, Peking University, Beijing, 100871, China.

[3] College of Chemistry and Molecular Engineering, Peking University, Beijing, 100871, China.

* Corresponding author. Email: doujinhu@pku.edu.cn (J. D.)

**ABSTRACT:** Conductive metal–organic frameworks (c-MOFs) are promising electronic materials, where electronic states critically determine electrical conductivity. However, achieving controllable modulation of these states during crystallization remains challenging. Existing approaches rely either on redesigning framework components or on introducing external dopants after crystallization, making electronic-state regulation dependent on structural modification or postsynthetic doping. Here, we demonstrate a crystallization-enabled in situ proton-management strategy for regulating the electronic state of c-MOFs. Using $Co_9HHTP_4$ as a model system, we show that distinct proton environments influence the balance between competing proton-coupled electron-transfer pathways during Co–O framework assembly while preserving the fundamental framework architecture. $HNO_3$-mediated crystallization produces an electronic state consistent with enhanced electron retention within the Co–O–π framework and results in a single-crystal conductivity of $8.4 \times 10^{-2}$ S cm$^{-1}$, nearly four orders of magnitude higher than $Co_9HHTP_4$-NaOAc ($1.7 \times 10^{-5}$ S cm$^{-1}$), together with a substantially reduced transport activation energy. Hall measurements further reveal a pronounced increase in carrier concentration for $Co_9HHTP_4$-$HNO_3$, providing direct evidence that the enhanced conductivity is associated with a substantially increased carrier population. These results establish crystallization conditions as an active parameter for programming electronic states and charge transport in conductive frameworks.

## Introduction

Two-dimensional conductive metal–organic frameworks (2D c-MOFs), constructed from metal ions and π-conjugated ligands[1], integrate electronic conductivity with the permanent porosity and chemical tunability of MOFs[2], enabling applications in catalysis[3], energy storage[4-5], sensing[6], and electronic devices[7]. However, the intrinsic electronic structures determined by the framework composition do not always provide optimal carrier populations for efficient charge transport, highlighting the need for further regulation of the electronic state of conductive frameworks[1-2, 8-9]. Consequently, strategies that enable precise regulation of electronic filling and carrier density provide important opportunities for optimizing the electrical properties of c-MOFs.

Existing strategies for regulating the electronic states of 2D c-MOFs can be broadly classified into pre-synthetic framework engineering and postsynthetic doping[10-12]. Pre-synthetic approaches focus on modulating framework composition and electronic structure, including the design of π-conjugated ligands, redox-active building blocks, and metal nodes with optimized electronic coupling[13-15]. Although effective, these approaches generally require modification of the framework components themselves, making electronic optimization inherently associated with changes in chemical composition and structural characteristics. Alternatively, postsynthetic doping strategies have been developed to further regulate carrier density and electrical conductivity in 2D c-MOFs[10]. For example, iodine-vapour treatment has been widely used to oxidatively dope preformed frameworks through charge transfer between $I_2$ and the electronically active lattice[16]. Similarly, redox-active molecular guests or conducting species can be introduced into the pores of an already crystallized MOF to modify its electronic filling[17]. Although effective, these approaches are predominantly postsynthetic, requiring dopant molecules to access a preformed porous lattice, where incorporation can be influenced by diffusion pathways and the availability of accessible binding sites[18-20]. These constraints can limit control over doping depth, spatial homogeneity and the local electronic state of the incorporated species, particularly in highly crystalline or large-crystal materials[18, 21] (Figure 1a). This challenge creates an inherent tension between preserving high crystallinity, which is desirable for intrinsic charge transport, and achieving homogeneous electronic modulation throughout the crystal. Unlike postsynthetic doping, crystallization provides a unique opportunity to regulate electronic states during framework assembly, when charge transfer, ligand oxidation, and coordination are intrinsically coupled. Establishing the electronic state during framework formation could therefore enable more homogeneous electronic-state regulation while avoiding the diffusion limitations inherent to postsynthetic guest incorporation.

Here, using $Co_9HHTP_4$ as a model 2D c-MOF, we demonstrate an *in situ* proton-management strategy in which the electronic state of the framework is regulated directly during crystallization (Figure 1b). $HNO_3$ and NaOAc establish distinct proton-management environments while still allowing the formation of well-crystallized single crystals. Importantly, the proton-rich crystallization environment created by $HNO_3$ produces a more electron-rich framework and markedly enhances electrical conductivity, with $Co_9HHTP_4$-$HNO_3$ reaching $8.4 \times 10^{-2}$ S cm$^{-1}$, representing an enhancement of nearly four orders of magnitude compared with $Co_9HHTP_4$-NaOAc ($1.7 \times 10^{-5}$ S cm$^{-1}$). Hall measurements directly reveal a pronounced increase in carrier population: the Hall-derived carrier concentration increases from $1.14 \times 10^{15}$ cm$^{-3}$ in $Co_9HHTP_4$-NaOAc to $4.17 \times 10^{17}$ cm$^{-3}$ in $Co_9HHTP_4$-$HNO_3$, establishing enhanced carrier population as a key consequence of the crystallization-enabled electronic-state regulation. These results establish crystallization-enabled *in situ* proton-management as an effective strategy for regulating carrier density and charge transport in 2D c-MOFs without relying on postsynthetic dopant incorporation.

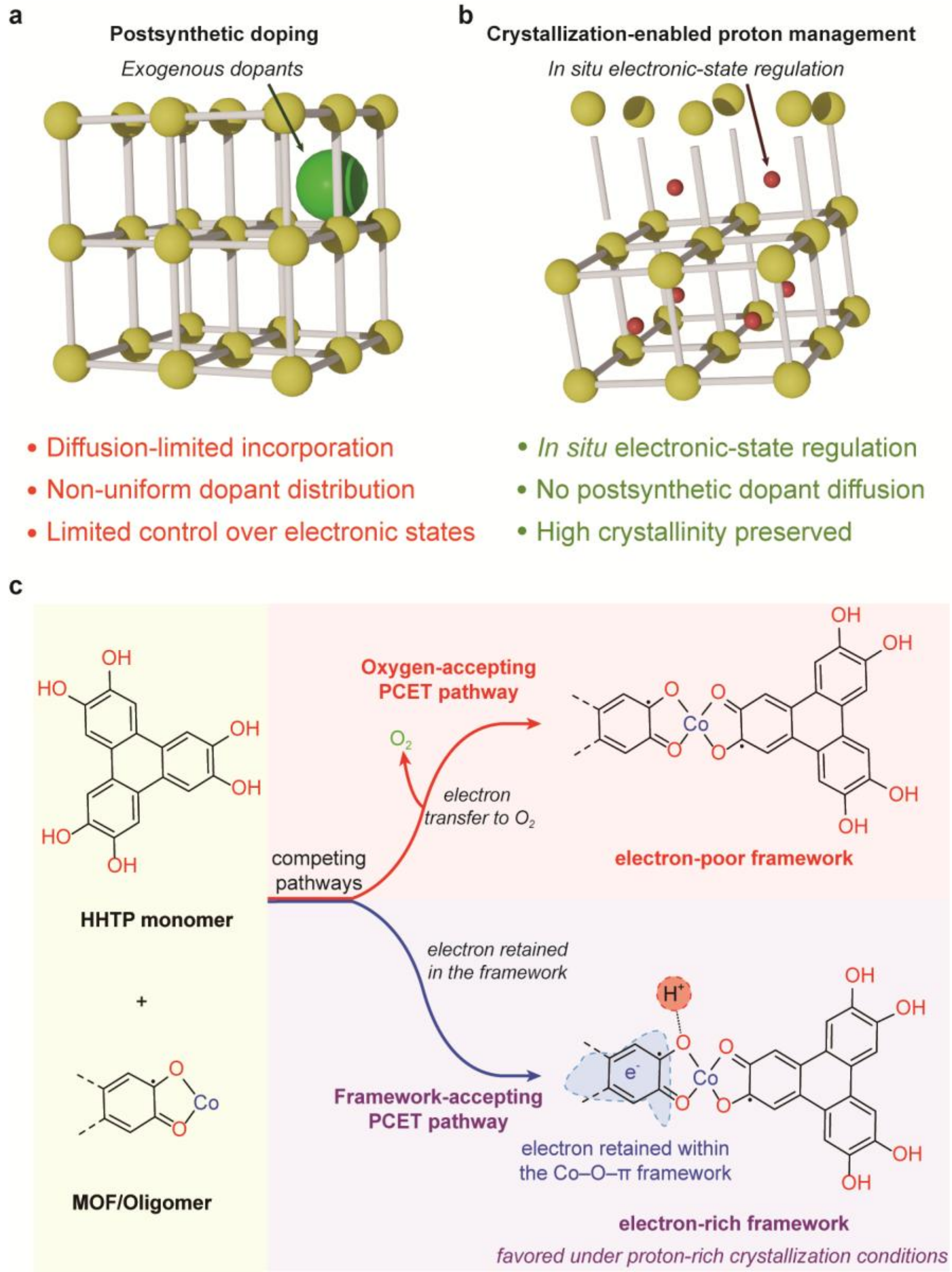

Figure 1. Schematic comparison of (a) conventional postsynthetic doping and (b) crystallization-enabled proton management. (c) The proposed oxygen-accepting and framework-accepting PCET pathways during $Co_9HHTP_4$ formation.

## Results and Discussion

### Proton-management-enabled electronic-state regulation

The formation of $Co_9HHTP_4$ involves the concurrent deprotonation and oxidation of HHTP together with Co–O coordination, making framework assembly intrinsically coupled to proton and electron transfer. These processes can therefore be described in terms of proton-coupled electron transfer (PCET). As previously proposed in our work, $Co_9HHTP_4$ formation may proceed through two competing PCET pathways[22] (Figure 1c). In the oxygen-accepting pathway, protons and electrons derived from HHTP are transferred to $O_2$, resulting in electron loss from the forming framework. By contrast, PCET involving preassembled $Co_9HHTP_4$ or oligomeric intermediates may favor electron retention within the Co–O–π network, leading to a more electron-rich framework. Because ligand deprotonation, oxidation, and metal coordination occur concurrently during framework assembly, proton activity may influence not only the kinetics of crystallization but also the relative energetics of the coupled proton- and electron-transfer processes. Under proton-rich conditions, suppressed ligand deprotonation may therefore alter the competition between electron transfer to $O_2$ and electron retention within partially assembled Co–O–π motifs, providing a mechanistic basis by which the crystallization environment could regulate the electronic state of the resulting framework.

We first established two distinct proton-management conditions for the growth of $Co_9HHTP_4$ single crystals. In the conventional NaOAc-mediated route, acetate facilitates HHTP deprotonation and Co–O coordination under stoichiometric $Co^{2+}$ conditions, thereby promoting framework formation and yielding $Co_9HHTP_4$-NaOAc. By contrast, introducing $HNO_3$ increases proton activity and creates a more acidic crystallization environment, which suppresses ligand deprotonation and slows Co–O framework formation. To compensate for this acid-induced kinetic inhibition, excess cobalt acetate was introduced to increase the availability of $Co^{2+}$ and sustain framework growth. This strategy enables the formation of large $Co_9HHTP_4$-$HNO_3$ single crystals under proton-rich conditions. The two synthetic routes therefore establish distinct crystallization regimes: acetate-assisted deprotonation favors framework formation under less acidic conditions, whereas acid-suppressed deprotonation combined with excess $Co^{2+}$ allows crystallization to proceed in a proton-rich environment. Detailed synthetic conditions are provided in the Supporting Information.

We next examined whether the two proton-management routes altered the crystalline framework of $Co_9HHTP_4$. The PXRD patterns of $Co_9HHTP_4$-$HNO_3$ and $Co_9HHTP_4$-NaOAc display the same set of characteristic reflections with closely matched peak positions, indicating that both products possess the same crystalline phase and comparable long-range order (Figure S1). Importantly, the single-crystal structure of $Co_9HHTP_4$-$HNO_3$ was resolved for the first time in this work. Side-by-side comparison with the previously reported $Co_9HHTP_4$-NaOAc structure[22] shows that both materials exhibit the same in-plane Co–HHTP connectivity and the same AB-type stacking sequence (Figure 2a–f). Their interlayer metrics are also closely matched, with corresponding separations of 3.59 and 3.12 Å for $Co_9HHTP_4$-$HNO_3$ and 3.57 and 3.12 Å for $Co_9HHTP_4$-NaOAc. The acidic crystallization route therefore enables the growth and structural determination of $Co_9HHTP_4$-$HNO_3$ single crystals while preserving the fundamental $Co_9HHTP_4$ framework topology and stacking architecture**.**

SEM images further show that both samples consist predominantly of micrometer-scale elongated rod-like crystals, although differences in crystal-size distribution and aggregation are evident (Figure 2g,h). High-resolution cryo-EM further reveals well-defined periodic lattice fringes in both samples (Figure 2i,j), with long-period repeats of approximately 1.97 and 1.89 nm for $Co_9HHTP_4$-$HNO_3$ and $Co_9HHTP_4$-NaOAc, respectively. Corresponding FFT analysis gives lattice spacings of approximately 3.91 and 3.75 Å. These observations further support highly ordered layered structures in both samples, while revealing modest differences in local lattice metrics.

Collectively, PXRD, cryo-EM and single-crystal diffraction show that the two samples retain closely comparable $Co_9HHTP_4$ framework architectures and high crystalline order. These results establish that proton-management conditions can regulate $Co_9HHTP_4$ crystallization while preserving the fundamental framework architecture, thereby providing a well-defined structural basis for evaluating how the distinct crystallization environments influence the electronic properties of the resulting materials.

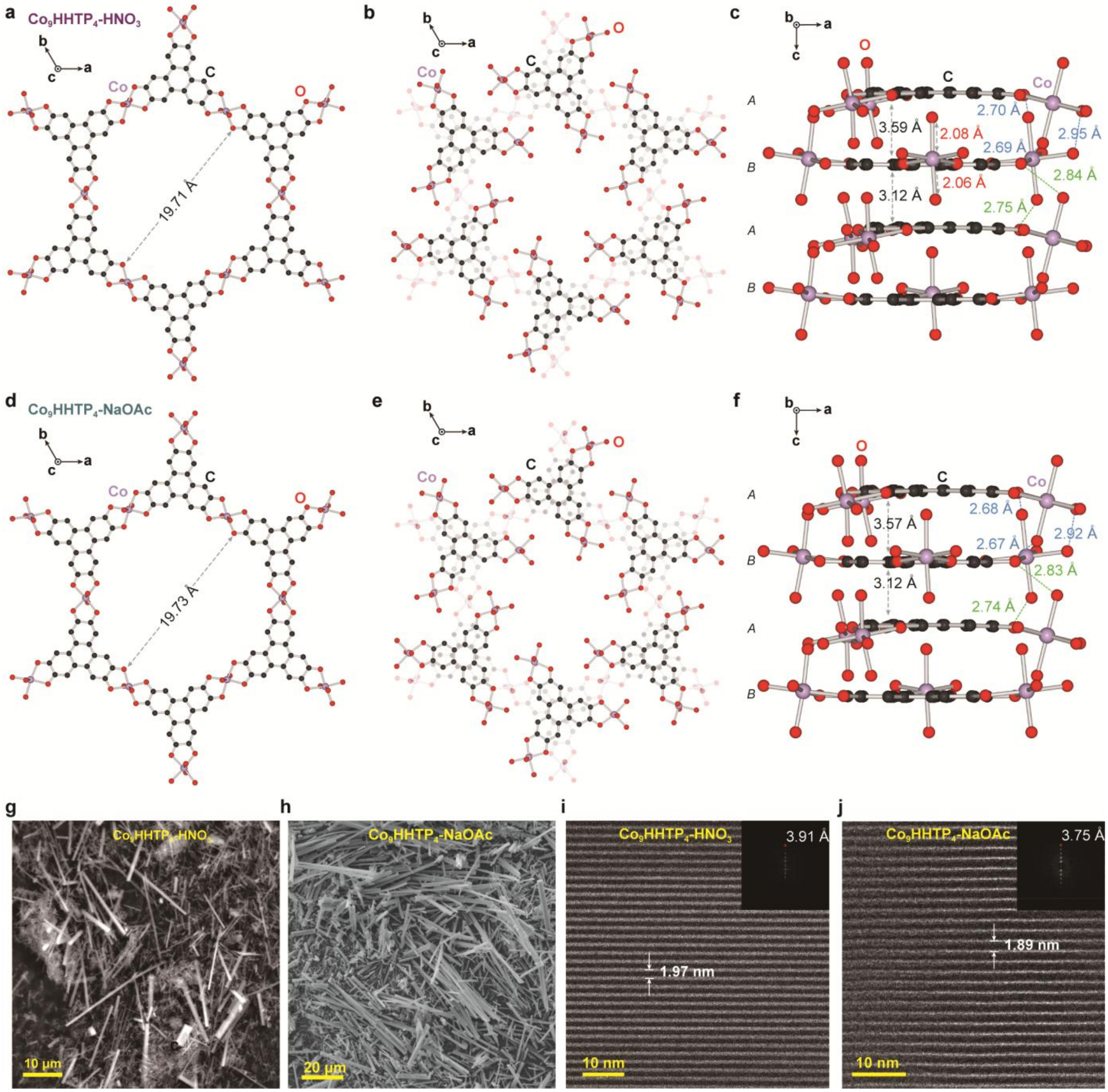

Figure 2. Structural comparison of $Co_9HHTP_4$ synthesized through distinct proton-management routes. (a–f) Single-crystal structures of $Co_9HHTP_4$-$HNO_3$ and $Co_9HHTP_4$-NaOAc derived from single-crystal X-ray diffraction. (a,b) and (d,e) show the single and double layers viewed along the c axis of $Co_9HHTP_4$-$HNO_3$ and $Co_9HHTP_4$-NaOAc, respectively, with (b) and (e) highlighting the rotated-eclipsed registry between the molecular clusters and extended sheets. (c) and (f) show views along the b axis, illustrating the (AB)n bilayer stacking motif in both structures. Interlayer hydrogen bonds are indicated by blue and green dashed lines. (g,h) SEM images showing elongated rod-like morphologies. (i,j) High-resolution cryo-EM images and corresponding FFT patterns, revealing periodic layered ordering in $Co_9HHTP_4$-$HNO_3$ and $Co_9HHTP_4$-NaOAc, respectively. The labeled values indicate long-period fringe spacings and FFT-derived lattice spacings.

**Distinct proton environments and electronic-state modulation**

Having established closely comparable $Co_9HHTP_4$ framework architectures, we next examined how the distinct proton-management conditions were manifested in the reaction environment and reflected in the chemical states of the resulting crystals. The two synthetic routes produced clearly distinct solution-phase proton environments, as revealed by their pH evolution (错误!未找到引用源。a). In the $HNO_3$-mediated route, the pH decreased slightly from approximately 5.70 to 5.45 during the initial stage and subsequently remained within a narrow range of approximately 5.45–5.56. This relatively small variation indicates that the reaction proceeded under persistently acidic conditions with comparatively stable proton activity. In contrast, the NaOAc-mediated solution exhibited a pronounced decrease in pH from approximately 6.50 to 6.00, followed by a more gradual decline to approximately 5.74. This progressive acidification is consistent with net proton release accompanying ligand deprotonation and framework formation, partially moderated by acetate. Thus, the $HNO_3$ route maintains a higher and more stable proton activity, whereas the NaOAc route undergoes more pronounced reaction-associated acidification. These contrasting pH profiles establish two distinct proton-management regimes during framework formation, which may alter the relative contributions of the two competing PCET pathways.

$N_2$ adsorption measurements show that $Co_9HHTP_4$-$HNO_3$ has a lower BET surface area of 288.47 $m^2$ $g^{-1}$ than $Co_9HHTP_4$-NaOAc, which exhibits a surface area of 333.67 $m^2$ $g^{-1}$ (Figure 3c). Given the closely comparable framework connectivity and stacking architectures, the lower accessible surface area of $Co_9HHTP_4$-$HNO_3$ may reflect differences in the internal pore environment, including possible proton-associated species. FTIR spectra provide complementary evidence for a distinct hydrogen-bonding environment. $Co_9HHTP_4$-$HNO_3$ exhibits a broader and more intense O–H stretching envelope in the 2500–3700 $cm^{-1}$ region than $Co_9HHTP_4$-NaOAc (Figure 3b), indicating a larger population of strongly hydrogen-bonded O–H-containing species. Unlike the oxygen-accepting pathway, which does not require additional cationic charge compensation within the framework, the framework-accepting pathway results in additional negative charge being retained within the Co–O–π network. In the context of the proposed framework-accepting PCET pathway, this proton-rich environment is consistent with the presence of proton-associated species that could compensate the additional negative charge retained within the Co–O–π network.To further probe the proton-associated species inferred from the FTIR measurements, proton-conductivity measurements were performed using Nafion-blocking configurations. At 25°C and 30% RH, $Co_9HHTP_4$-$HNO_3$ exhibits a proton conductivity of $3.79\times10^{-6}$ S $cm^{-1}$ (Figure 3d and Figure S7), approximately five times higher than that of $Co_9HHTP_4$-NaOAc ($7.06\times10^{-7}$ S $cm^{-1}$). The enhanced proton conductivity provides further evidence for a more pronounced proton-associated transport environment in $Co_9HHTP_4$-$HNO_3$, consistent with the proposed proton-associated charge-compensation picture. Importantly, the proton conductivity remains far below the measured electronic conductivity, indicating that proton transport itself is not the dominant origin of the enhanced electrical conductivity.

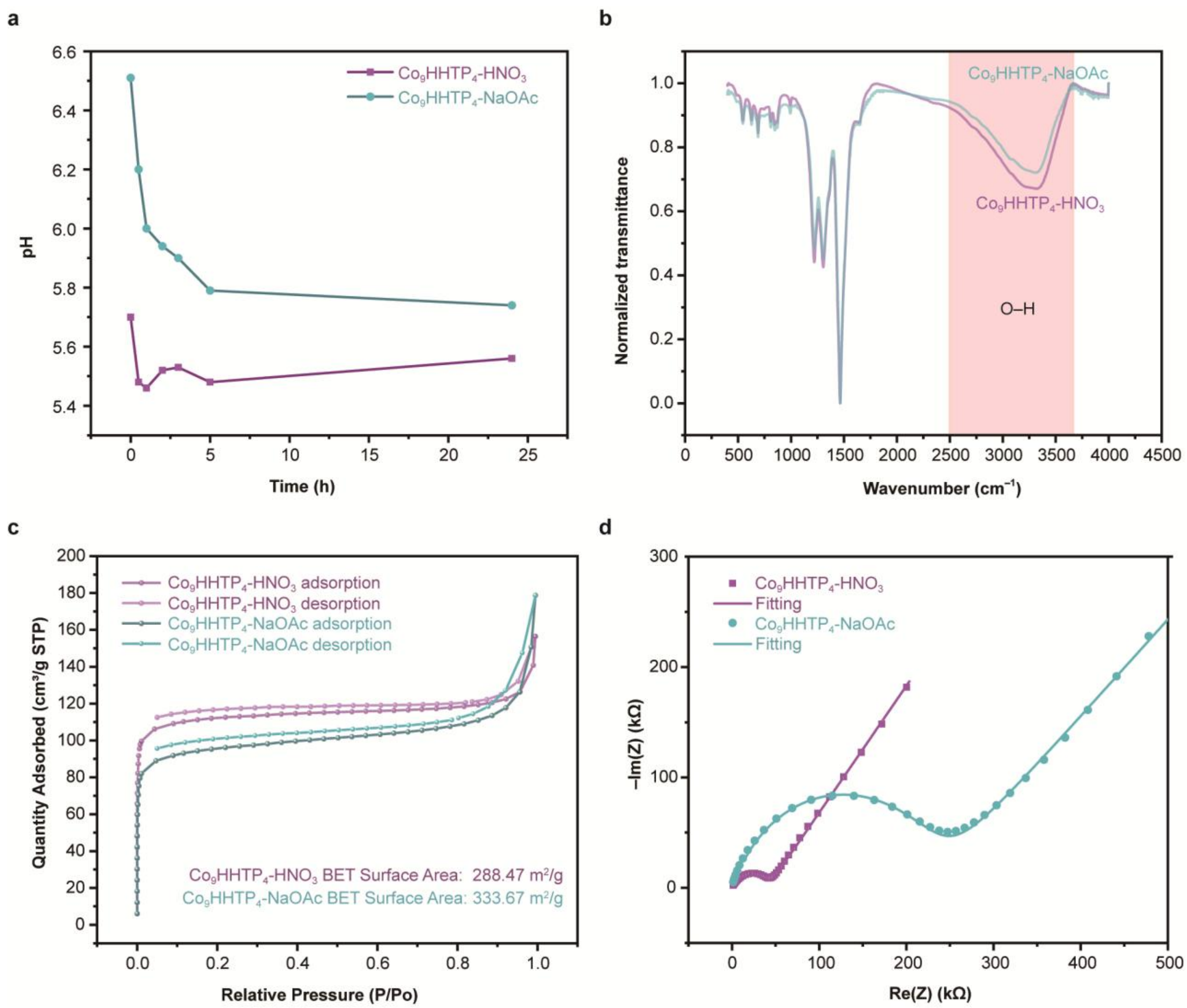

Figure 3. Distinct proton-management environments in $Co_9HHTP_4$ synthesized under different crystallization conditions. (a) pH evolution of the precursor solutions during $Co_9HHTP_4$ formation, showing different proton activities in $HNO_3$- and NaOAc-mediated routes. (b) FTIR spectra highlighting the enhanced O–H stretching features of $Co_9HHTP_4$-$HNO_3$. (c) $N_2$ adsorption–desorption isotherms and BET surface areas of $Co_9HHTP_4$-$HNO_3$ and $Co_9HHTP_4$-NaOAc. (d) proton-conductivity measurements of $Co_9HHTP_4$-$HNO_3$ and $Co_9HHTP_4$-NaOAc derived from electrochemical impedance spectroscopy.

X-ray photoelectron spectroscopy was further used to determine whether the $HNO_3$-mediated crystallization route induced a substantial change in the oxidation state of the Co centres. The Co 2p spectra of $Co_9HHTP_4$-$HNO_3$ and $Co_9HHTP_4$-NaOAc exhibit closely comparable Co $2p_{3/2}$ and Co $2p_{1/2}$ peak positions, together with similar shake-up satellite features (Figure S3 and S4), providing no clear evidence for a substantial difference in the average Co oxidation state between the two samples. These results suggest that the electronic modification induced under proton-rich crystallization conditions is unlikely to originate primarily from Co-centred redox chemistry. Instead, they are more consistent with a redistribution of electronic charge within the extended Co–O–π framework. Because DMF is used during synthesis, the N 1s region cannot by itself be used to assign nitrate-derived species unless a distinct high-binding-energy nitrate component can be resolved. The XPS analysis is therefore used primarily to argue against a major Co-centred redox change rather than to identify the protonic charge-compensating species. Diffuse-reflectance UV–vis measurements further show nearly identical optical gaps of approximately 0.85 and 0.84 eV for $Co_9HHTP_4$-$HNO_3$ and $Co_9HHTP_4$-NaOAc, respectively (Figure S5). The close agreement indicates that the two crystallization routes do not substantially alter the fundamental optical gap of the $Co_9HHTP_4$ framework.

Taken together, these measurements establish a consistent picture in which the two crystallization routes generate distinct proton-associated environments while preserving closely comparable framework structures. The pH evolution directly demonstrates different proton activities during framework formation, whereas the enhanced O–H stretching features and higher proton conductivity of $Co_9HHTP_4$-$HNO_3$ further support a more pronounced hydrogen-bonded and proton-associated environment in the resulting material. The modest reduction in BET surface area is also consistent with differences in the internal pore environment, although it does not by itself establish the amount or location of proton-associated species. At the same time, the closely comparable Co 2p spectra provide no evidence for a substantial change in the average Co oxidation state, and the nearly identical optical gaps indicate that the electronic differences are unlikely to originate from a major reconstruction of the framework electronic structure. These observations therefore narrow the likely origin of the electronic modulation toward differences in the electronic filling of the extended Co–O–π network rather than changes in framework architecture, Co-centred oxidation state, or optical gap. Within the proposed PCET picture, the proton-rich crystallization environment may favor greater electron retention within the framework, with proton-associated species providing charge compensation. Direct transport measurements are therefore required to determine whether this proposed electronic-state modulation is reflected in the carrier population.

**Carrier-density-driven electrical transport enhancement**

To exclude possible artifacts introduced by electron beam lithography (EBL) based device fabrication, we additionally fabricated direct two-terminal contacts using carbon paste and examined the intrinsic transport trend of the two distinct $Co_9HHTP_4$ crystals. The two samples exhibit markedly different electrical transport properties. $Co_9HHTP_4$-$HNO_3$ shows a single-crystal conductivity of $8.4 \times 10^{-2}$ S $cm^{-1}$ (Figure 4a, Figure S8 and S10) and a relatively low activation energy of 0.18 eV (Figure 4b), whereas $Co_9HHTP_4$-NaOAc exhibits a much lower conductivity of $1.7 \times 10^{-5}$ S $cm^{-1}$ (Figure S11) and a higher activation energy of 0.67 eV (Figure 4c). The latter is in reasonable agreement with the previously reported value of 0.48 eV[22]. These results demonstrate that the $HNO_3$-mediated crystallization route markedly enhances charge transport and substantially lowers the energetic barrier for carrier motion.

Bulk powder measurements reproduce the same conductivity trend. $Co_9HHTP_4$-$HNO_3$ and $Co_9HHTP_4$-NaOAc exhibit powder conductivities of $1.3\times10^{-2}$ and $2.8\times10^{-4}$ S $cm^{-1}$, respectively (Figure 4d and Figure S12), corresponding to an approximately 46-fold difference. The higher powder conductivity of $Co_9HHTP_4$-NaOAc compared with its single-crystal value may arise from orientation-dependent charge transport in the strongly anisotropic layered framework, together with differences in measurement geometry and interparticle conduction pathways. We therefore focus primarily on the robust conductivity contrast between the two materials rather than on the absolute difference between single-crystal and powder values. Notably, for both materials, the proton conductivity remains substantially lower than the corresponding bulk electrical conductivity, indicating that proton transport is unlikely to be the dominant contributor to the observed electrical transport.

Hall measurements were further performed to quantitatively distinguish the contributions of carrier concentration and mobility to the conductivity enhancement (**错误!未找到引用源。**e). $Co_9HHTP_4$-$HNO_3$ exhibits a Hall-derived carrier concentration of $4.17 \times 10^{17}$ $cm^{-3}$, approximately 366 times higher than that of $Co_9HHTP_4$-NaOAc ($1.14 \times 10^{15}$ $cm^{-3}$). This pronounced increase provides direct evidence for a substantially increased carrier population and is consistent with altered electronic filling of the Co–O–π framework.

Importantly, the Hall mobility does not exhibit a corresponding enhancement; instead, it is lower in $Co_9HHTP_4$-$HNO_3$ ($2.77 \times 10^{-2}$ $cm^2$ $V^{-1}$ $s^{-1}$) than in $Co_9HHTP_4$-NaOAc ($1.52 \times 10^{-1}$ $cm^2$ $V^{-1}$ $s^{-1}$). Thus, the enhanced bulk conductivity cannot be attributed to improved carrier mobility and is predominantly associated with the substantially increased carrier concentration. Consistent with $\boldsymbol{\sigma = ne\mu}$, the Hall-derived carrier concentration and mobility correspond to an approximately 67-fold difference in conductivity between the two samples, in reasonable agreement with the approximately 46-fold conductivity enhancement independently observed in the bulk powder measurements.

These Hall results provide direct transport evidence that the distinct crystallization environments produce substantially different carrier populations despite their closely comparable framework structures and optical gaps. Together with the pH, FTIR, proton-conductivity and XPS results, the increased carrier concentration in $Co_9HHTP_4$-$HNO_3$ supports the proposed picture in which proton-rich crystallization favors greater electron retention within the Co–O–π framework, accompanied by proton-associated charge compensation.

Having experimentally established a substantially increased carrier population in $Co_9HHTP_4$-$HNO_3$, we next examined how proton association influences the electronic coupling relevant to charge transport. Representative interlayer dimer models were extracted from the experimentally determined crystal structure, and an additional proton was introduced to represent the proposed proton-associated, electron-retaining state. Because interlayer coupling constitutes an important component of charge transport in stacked 2D c-MOFs, the calculated transfer integral provides a qualitative measure of the electronic coupling between adjacent layers rather than a direct simulation of the full transport process. The calculated electronic transfer integral ($t_e$) decreases moderately from 299.3 meV in the unprotonated model to 241.1 meV after proton incorporation (错误!未找到引用源。f), indicating a slight weakening of the interlayer electronic coupling. This result is consistent with the Hall measurements, which show a decrease in carrier mobility for $Co_9HHTP_4$-$HNO_3$. Together, the calculations and Hall measurements indicate that the conductivity enhancement is dominated by the substantially increased carrier concentration rather than stronger electronic coupling or higher carrier mobility. Within the proposed PCET framework, these results support a picture in which proton-rich crystallization favors electron retention and increased electronic filling of the Co–O–π framework, thereby enhancing electrical conductivity primarily through carrier-density regulation.

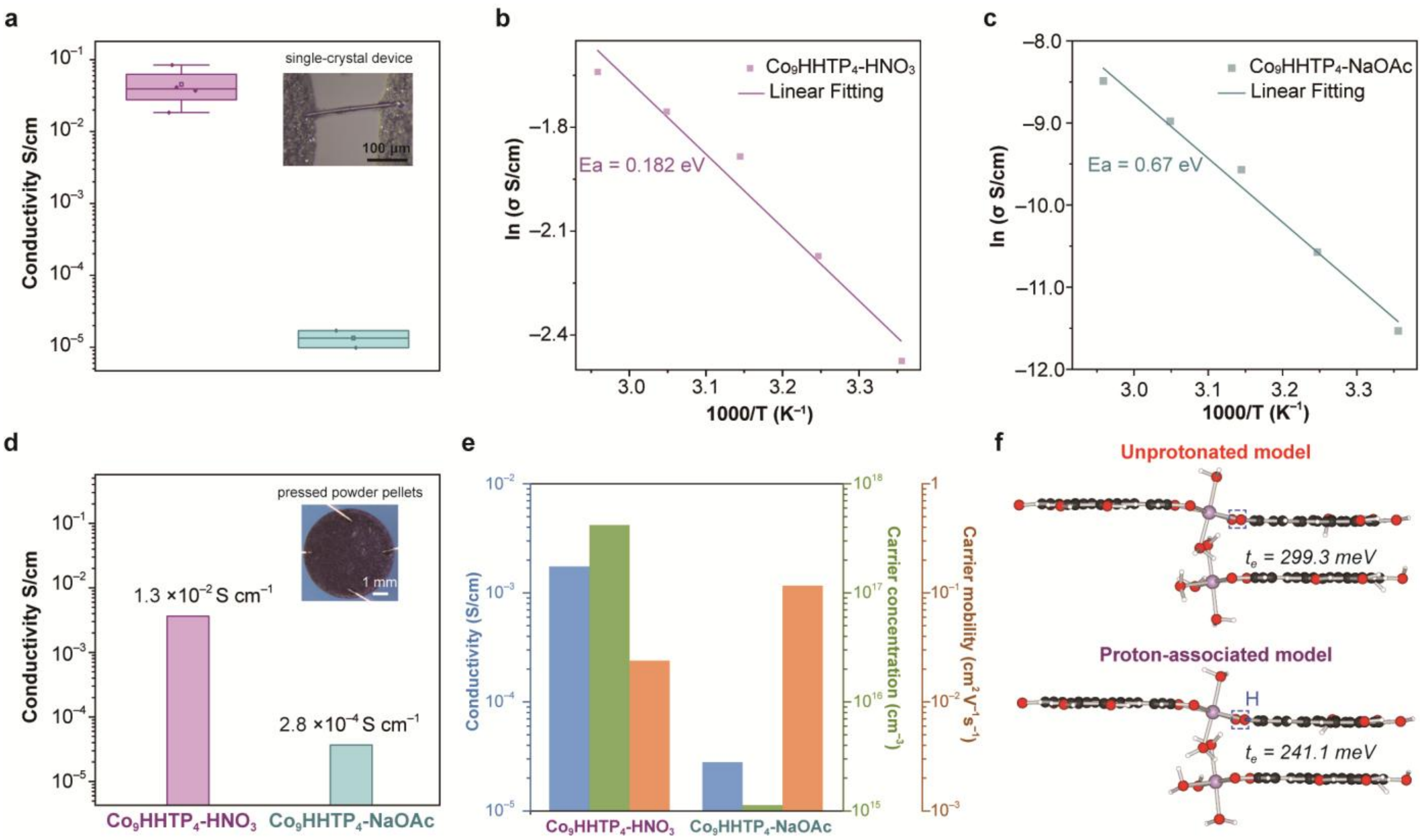

Figure 4. Electrical transport characteristics and charge-transport analysis of $Co_9HHTP_4$ synthesized under different proton-management conditions. (a) Single-crystal electrical conductivity of $Co_9HHTP_4$-$HNO_3$ and $Co_9HHTP_4$-NaOAc measured using direct two-terminal contacts. Inset: representative optical image of a contacted single crystal. (b,c) Arrhenius plots of the temperature-dependent conductivity for $Co_9HHTP_4$-$HNO_3$ and $Co_9HHTP_4$-NaOAc, respectively, with the corresponding activation energies obtained from linear fitting. (d) Bulk powder conductivity of $Co_9HHTP_4$-$HNO_3$ and $Co_9HHTP_4$-NaOAc. Inset: representative image of the pressed pellet used for electrical measurements. (e) Carrier-density-driven conductivity enhancement revealed by Hall measurements. (f) Calculated interlayer electronic transfer integrals ($t_e$) for unprotonated and proton-associated $Co_9HHTP_4$ models. Proton incorporation slightly decreases the interlayer electronic coupling from 299.3 to 241.1 meV.

## Conclusion

In summary, we demonstrate that the electronic state and charge transport of $Co_9HHTP_4$ can be regulated directly through the crystallization environment without altering the fundamental framework architecture. $HNO_3$-mediated crystallization establishes a proton-rich environment that is consistent with altered PCET pathways and enhanced electron retention within the Co–O–π framework. The resulting $Co_9HHTP_4$-$HNO_3$ exhibits an enhancement of nearly four orders of magnitude in single-crystal conductivity and a substantially reduced transport activation energy compared with $Co_9HHTP_4$-NaOAc. Hall measurements further reveal an approximately 366-fold increase in carrier concentration, directly establishing the substantially increased carrier population as the dominant origin of the enhanced bulk conductivity. Together with the structural, spectroscopic, and proton-transport results, these findings support a picture in which proton-rich crystallization favors greater electron retention and electronic filling within the Co–O–π framework. More broadly, these results identify crystallization conditions as a previously underexplored degree of freedom for programming electronic states and charge transport in c-MOFs. This concept may be particularly relevant to redox-active frameworks in which ligand oxidation, proton transfer, and metal coordination are intrinsically coupled during framework formation, offering a potential route to regulate carrier density without redesigning the framework components or relying on postsynthetic dopant diffusion.

## ASSOCIATED CONTENT

### Supporting Information

The following files are available free of charge.
Complete experimental procedures and other supporting data. (PDF)
The crystallographic data have been deposited at the Cambridge Crystallographic Data Centre (CCDC) with accession numbers CCDC 2591648 (for $Co_9HHTP_4$-$HNO_3$) (CIF).

## AUTHOR INFORMATION

### Corresponding Author

*Jin-Hu Dou
doujinhu@pku.edu.cn (J. D.)

### Funding Sources

This work was financially supported by the National Key R&D Program of China (Grant No. 2023YFE0206400), National Natural Science Foundation of China (Grant No. 22171185, 22575005), and the Clinical Medicine Plus X-Young Scholars Project at Peking University backed by the Fundamental Research Funds for the Central Universities.

### Notes

The authors declare no competing interests.

### ACKNOWLEDGMENT

This work was financially supported by the National Key R&D Program of China (Grant No. 2023YFE0206400), National Natural Science Foundation of China (Grant No. 22171185, 22575005), and the Clinical Medicine Plus X-Young Scholars Project at Peking University backed by the Fundamental Research Funds for the Central Universities. The support of the beamline staff at BL14B1 and BL17B of SSRF is acknowledged for their assistance with data acquisition and analysis. Cryo-EM was performed at Electron Microscopy Laboratory of Peking University with the assistance of Xuemei Li. We appreciate the Molecular Materials and Nanofabrication Laboratory (MMNL) in the College of Chemistry and Electron Microscopy Laboratory of Peking University for the support of instruments to perform device fabrication and characterization. We sincerely appreciate Shucui Sun and Quantum Design China for helpful discussions in the transport measurement. We acknowledge the High-Performance Computing Platform of Peking University for supporting the computational work.

SYNOPSIS TOC

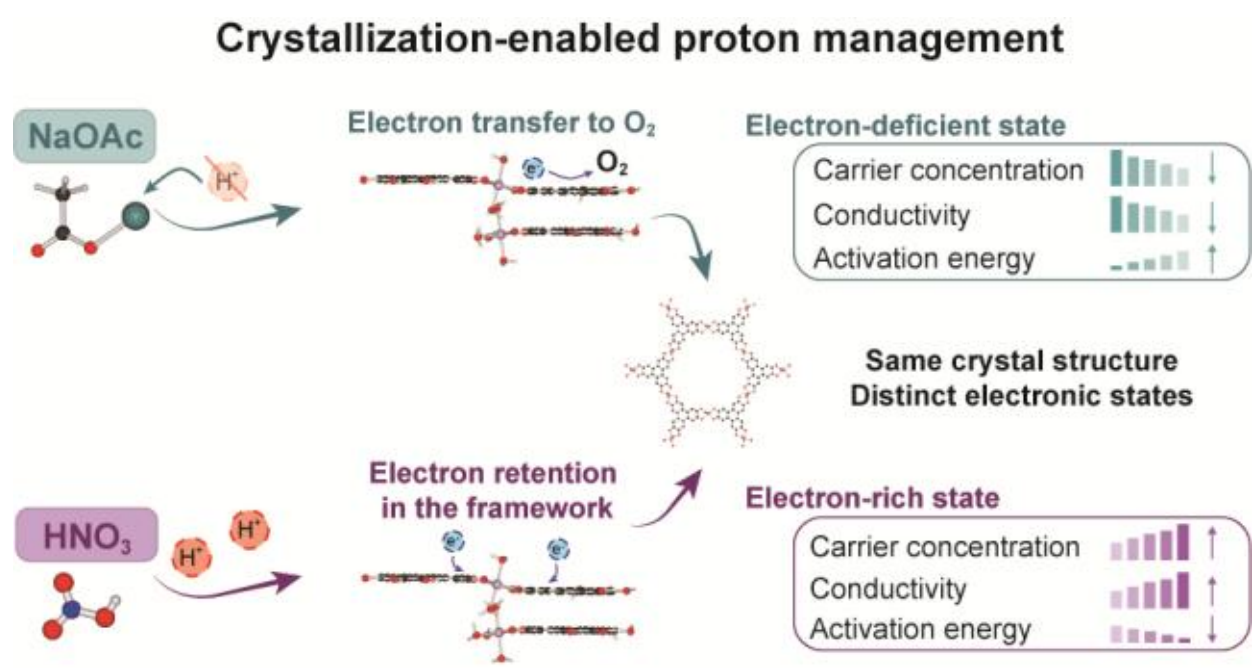